\documentclass [12pt]{article}

\begin{document}

\begin{center}
\textbf{\huge Quantum spectrum and statistic entropy of black
hole}
\end{center}

\begin{center}
Zhao Ren$^{a,b,}$\footnote{ E-mail address:
zhaoren2969@yahoo.com.cn }, Li Huai-Fan$^a$ and Zhang Sheng-Li$^b$

$^a$Department of Physics, Shanxi Datong University, Datong 037009
P.R.China\\ $^b$Department of Applied Physics, Xi'an Jiaotong
University, Xi'an 710049 P.R.China

\end{center}

\begin{center}
\textbf{\Large Abtract}
\end{center}

Taking the horizon surface of the black hole as a compact membrane
and solving the oscillation equation of this membrane by
Klein-Gordon equation, we derive the frequencies of oscillation
modes of the horizon surface, which are proportional to the
radiation temperature of the black hole. However, the frequencies
of oscillation modes are not equidistant. Using the distribution
of obtained frequencies of oscillation mode we compute the
statistic entropy of the black hole and obtain that the statistic
entropy of the black hole is proportional to the area of the
horizon. Therefore, it is proven that the quantum statistic
entropy of the black hole is consistent with Bekenstein-Hawking
entropy.
\vspace{1.cm}\\

The necessity for a quantum theory of gravity was already
recognized in the 1930s. However, despite the flurry of activity
on this subject we still lack a complete theory of quantum
gravity. It is believed that black holes may play a major role in
our attempts to shed some light on the nature of a quantum theory
of gravity (such as the role played by atoms in the early
development of quantum mechanics) [1].

Quantization of the black hole horizon area is a long-standing
problem. Since it has been proposed in 1974 by Bekenstein, this
problem is entirely unsolved [2]. It involves the quantum origin
of the black hole entropy. After Hod derived the interval between
quantum spectrums of Schwarzschild black hole using Bohr
correspondence principle, considerable progress in this field had
been done in the past few years [3-6]. However, the interval of
spectrums given by Refs. [1, 7] are equidistant. For complex
space-time, we only obtain the numerical solution to frequency.

A basis for the Hilbert space of loop gravity is given by spin
networks. These are graphs whose edges are labeled by
representations of the gauge group of the theory. In the case of
gravity this group is taken to be SU(\ref{eq2}) and the
representations are thus labeled by positive half-integers $j =
0,1 / 2,1,3 / 2, \cdots $. If a surface is intersected by an edge
of such a spin network carrying the label $j$ the surface acquires
the area [8,9]

\begin{equation}
\label{eq1}
A(j) = 8\pi l_p^2 \gamma \sqrt {j(j + 1)} ,
\end{equation}

\noindent where $l_p $ is the Planck length and $\gamma $ is the
so-called Immirzi parameter. Ref.[9] has
given the value of
parameter $\gamma $.

In this paper, taking the horizon surface of the black hole as a
compact membrane and supposing that on the membrane the amplitude
of wave is much smaller than the radius of horizon, we constitute
the wave equation describing the propagation of surface wave on
horizon surface. Comparing this equation with the angular equation
of Klein-Gordon equation followed by the massless particles in the
curved space-time, we can obtain the vibration frequency of the
surface wave propagation on the black hole surface. This frequency
is not equidistant and its quantum characteristic is similar to
the quantum property of the black hole area given by
Eq.(\ref{eq1}).To simplify the discussion, we take $c = \hbar =
l_p = K_B $.

The line element of the Schwarzschild black hole is

\begin{equation}
\label{eq2}
ds^2 = - A(r)dt^2 + A^{ - 1}(r)dr^2 + r^2(d\theta ^2 + \sin ^2\theta
d\varphi ^2),
\end{equation}

\noindent where $A(r) = 1 - 2M / r$, $M$ is the mass of the black
hole.

In curved space-time, Klein-Gordon equation followed by massless particles
is

\begin{equation}
\label{eq3}
\frac{1}{\sqrt { - g} }\frac{\partial }{\partial x^\mu }\left( {\sqrt { - g}
g^{\mu \nu }\frac{\partial }{\partial x^\nu }} \right)\Psi = 0.
\end{equation}
Substituting the metric (\ref{eq2}) into Eq.(\ref{eq3}) and
separating the variables based on the symmetry of the space-time,
we have

\begin{equation}
\label{eq4}
\Psi (t,r,\theta ,\varphi ) = f(r)\chi (\theta )\Phi (\varphi )\exp ( - iEt)
= f(r)\chi (\theta )\exp ( - i(Et - m\varphi )).
\end{equation}
The radial and angular components satisfy the following equation
[10, 11]

\begin{equation}
\label{eq5}
\frac{E^2}{A(r)}f(r) + \frac{1}{r^2}\frac{d}{dr}\left[
{r^2A(r)\frac{df(r)}{dr}} \right] - \frac{l(l + 1)}{r^2}f(r) = 0,
\end{equation}

\begin{equation}
\label{eq6}
\frac{1}{\sin \theta }\frac{d}{d\theta }\left[ {\sin \theta \frac{d\chi
(\theta )}{d\theta }} \right] + \left( {(l(l + 1) - \frac{m^2}{\sin ^2\theta
}} \right)\chi (\theta ) = 0,
\end{equation}

\noindent
where $l = 0,1,2,3 \cdots $, and $l \ge \left| m \right|$, $E$ is the energy
of the particle, $m$ is a constant.

If we take the horizon surface of the black hole as a compact
spherical membrane, we can discuss Eq.(\ref{eq3}) and obtain the
wave equation of the surface wave propagation on the spherical
surface. To avoid divergent term in the equation, firstly, we
calculate the wave equation of the surface wave propagation near
horizon on $R = r_H + \varsigma $ the spherical surface ($r_H $ is
the location of the black hole horizon, $\varsigma $ is a positive
small quantity). Secondly, letting $\varsigma \to 0$, we derive
the wave equation of the surface wave propagation on the black
hole horizon surface. When $R$ is invariant, Eq.(\ref{eq3}) can be
reduced to

\begin{equation}
\label{eq7} \frac{1}{R^2\sin \theta }\frac{\partial }{\partial
\theta }\left( {\sin \theta \frac{\partial \psi (\theta ,\varphi
,t)}{\partial \theta }} \right) + \frac{1}{R^2\sin ^2\theta
}\frac{\partial ^2\psi (\theta ,\varphi ,t)}{\partial \varphi ^2}
= \frac{1}{A(R)}\frac{\partial ^2\psi (\theta ,\varphi
,t)}{\partial t^2},
\end{equation}
where the field function $\psi (\theta ,\varphi ,t)$ must be
single-valued and continuous everywhere. So the field function
satisfies the following conditions.

\begin{equation}
\label{eq8}
\chi (\theta ) = \chi (\theta + 2\pi ),
\quad
\Phi (\varphi ) = \Phi (\varphi + 2\pi ).
\end{equation}
using Separation of Variables ,we obtain

\begin{equation}
\label{eq9}
\psi (\theta ,\varphi ,t) = \chi (\theta )\Phi (\varphi )T(t),
\end{equation}
so

\begin{equation}
\label{eq10}
T(t) = B\sin (2\pi \nu t + \delta ),
\end{equation}

\noindent where $\delta $ is a phase constant, $\nu $ is a
variation frequency. Eq.(\ref{eq7}) can be rewritten as

\begin{equation}
\label{eq11}
\frac{1}{R^2\sin \theta }\frac{\partial }{\partial \theta }\left( {\sin
\theta \frac{\partial \chi (\theta )}{\partial \theta }} \right) + \left(
{\frac{4\pi ^2\nu ^2}{A(R)} - \frac{m^2}{R^2\sin ^2\theta }} \right)\chi
(\theta ) = 0.
\end{equation}
at any point the field function satisfies the Eq.(\ref{eq6}) on
the background of the black hole, the field function satisfies
Eq.(\ref{eq11}). So When in Eqs.(\ref{eq6}) and (\ref{eq11}) we
let $r = R$, they should be equivalence. Comparing Eq.(\ref{eq6})
with Eq.(\ref{eq11}), we obtain

\begin{equation}
\label{eq12}
l(l + 1) = \frac{4\pi ^2\nu ^2R^2}{A(R)},
\end{equation}
Since equation (6) should be consistent with equation (11),we
obtain a group of frequencies of oscillation modes, that is

\begin{equation}
\label{eq13}
\nu _l = \frac{A^{1 / 2}(R)}{2\pi R}\sqrt {l(l + 1)} .
\end{equation}
The frequency shift of the photon that is from the surface of the
star got by the observer at rest at an infinite distance is as
follows,

\begin{equation}
\label{eq14}
\nu = \nu ^0A^{1 / 2}(r),
\end{equation}

\noindent
where $\nu ^0$ is the natural vibration frequency on the surface of the
star, $\nu $ is the natural frequency of this particle measured by the
observer at rest at an infinite distance. Substituting (\ref{eq14}) into (\ref{eq13}) and
letting $R \to r_H $, we derive that the natural frequency of the black hole
horizon surface is

\begin{equation}
\label{eq15}
\nu _l^0 = \frac{1}{2\pi r_H }\sqrt {l(l + 1)} .
\end{equation}
Since $\left| m \right|$ can take value from 1 to $l$, and every
$m$ is double degenerate, if we add the mode $m = 0$, every
oscillation mode has $(2l + 1)$ degenerate. In addition, the
partial tone frequency is not the times of the lowest frequency
$\nu _1^0 = \frac{\sqrt 2 }{2\pi r_H }$ corresponding $l = 1$.
Mode frequency of $l = 0$ is zero. It represents the spherically
symmetric mode under the case of non-vibration.

Introducing partition function $Z$

\begin{equation}
\label{eq16}
Z = \sum\limits_l {(2l + 1)e^{ - \textstyle{\beta \over {2\pi r_H }}\sqrt
{l(l + 1)} }} ,
\end{equation}

\noindent where $\beta = 1 / T_H$ is the inverse Hawking radiation
temperature of the black hole. Taking $l$ as continuous
distribution and calculating (\ref{eq16}), we have

\begin{equation}
\label{eq17}
Z \approx \int\limits_0^\infty {2xe^{ - \textstyle{\beta \over {2\pi r_H
}}x}dx}
 = \frac{2(2\pi r_H )^2}{\beta ^2}.
\end{equation}
The entropy of the black hole is

\begin{equation}
\label{eq18}
S = N\left( {\ln Z - \beta \frac{\partial \ln Z}{\partial \beta }} \right)
 = N\left( {2 + \ln \frac{2(2\pi r_H )^2}{\beta ^2}} \right),
\end{equation}

\noindent where $N$ is the number of the particles. The internal
energy of the black hole is

\begin{equation}
\label{eq19}
U = M = - N\frac{\partial \ln Z}{\partial \beta } = \frac{2N}{\beta }.
\end{equation}
Substituting the radiation temperature of the black hole $T_H = 1
/ (4\pi r_H )$ and (\ref{eq19}) into (\ref{eq18}), we derive that
the entropy of the black hole is

\begin{equation}
\label{eq20}
S = \frac{A}{4}(2 - \ln 2)
 = \frac{A}{4}\times 1.3,
\end{equation}
where $A$ is the horizon area of the black hole. As a result, we
derive the relation between the quantum statistical entropy and
the horizon area. The difference between the result of
Eq.(\ref{eq20}) and Bekenstein-Hawking entropy is caused by the
fact that the integral replaces the sum in the calculation of the
partition function. The accurate result should be derived by
taking the sum of partition function.

According to Ref.[9], let the particles be at $l = 1$ quantum state. We have

\begin{equation}
\label{eq21}
Z = 3e^{ - \textstyle{\beta \over {2\pi r_H }}\sqrt 2 }.
\end{equation}
The entropy of the black hole is

\begin{equation}
\label{eq22}
S_{l = 1} = \frac{A}{4}\frac{\ln 3}{\sqrt 2 }.
\end{equation}
Because of $\frac{\ln 3}{\sqrt 2 } < 1$, $S_{l = 1} < A / 4$. That
is the lower limit of the black hole entropy is
$\frac{A}{4}\frac{\ln 3}{\sqrt 2 }$.

From (\ref{eq1}), we have

\begin{equation}
\label{eq23}
\Delta A = A(j) = 8\pi \gamma \sqrt {j(j + 1)} .
\end{equation}
Since the area $A$ and the $M$ of a Schwarzschild black hole are
related by

\begin{equation}
\label{eq24}
A = 16\pi M^2,
\end{equation}
Comparing (\ref{eq23}) with (\ref{eq24}), we obtain

\begin{equation}
\label{eq25}
\Delta M = \frac{\gamma }{4M}\sqrt {j(j + 1)} .
\end{equation}
Comparing (\ref{eq25}) with (\ref{eq15}), we derive $\gamma = 1 /
\pi $. As a result, we derive that the natural vibration frequency
of the black hole horizon surface (\ref{eq15}) is equivalence with
the spectrum of the black hole area given by (\ref{eq1}). However,
in our calculation there is not any uncertain factor.

Studying the radiation spectrum of the black hole is a very
interesting subject. However, until recently, the energy of the
radiation particles is derived by numerical calculations. For
rotating black hole, the calculation is more complex. In this
paper, taking the horizon surface of the black hole as a compact
membrane and solving the oscillation equation of this membrane, we
derive the frequencies of oscillation modes of the horizon
surface. Then using the derived frequencies, we compute the black
hole entropy. There is only small difference between our result
and Bekenstein-Hawking entropy. This departure is caused mostly by
the fact that the sum in Eq.(\ref{eq16}) is turned to the integral
in Eq.(\ref{eq17}). We obtain the statistical entropy of the black
hole under the case that there is not any assumption. Our result
is very close to Bekenstein-Hawking entropy which shows that the
quantum statistical entropy is consistent with the thermodynamic
entropy. In our paper, we derive the analytical solution of the
vibration frequency and in addition there is not any uncertain
parameter. We provide a new method for studying the quantum
characteristic of rotating black hole as well as
more complex black hole.\\
ACKNOWLEDGMENT

Zhao R acknowledges the help of Elias C.Vagenas,This project was
supported by the National Natural Science Foundation of China
under Grant No. 10374075 and the Shanxi Natural Science Foundation
of China under Grant No. 2006011012.

\textbf{REFERENCES}

[1] S. Hod, Phys.Rev. Lett. \textbf{81}, 4293 (1998).

[2] J. D. Bekenstein, Lett. Nuovo Cimento \textbf{11}, 467 (1974).

[3] V. Cordoso, J. P. S. Lemos, and S. Yoshida, Phys. Rev.
\textbf{D 69}, 044004(2004).

[4] S. Hod, Phys. Rev. \textbf{D 67}, 08150(R) (2003).

[5] C. Kunstatter, Phys. Rev. Lett. \textbf{90}, 161301 (2003).

[6] T. R. Choudhury, and T. Padmanabhan, Phys. Rev. \textbf{D 69},
064033 (2004).

[7] Jiliang Jing, Phys. Rev. \textbf{D 71}, 124006 (2005).

[8] C. Rovelli and L. Smolin, Nucl. Phys. \textbf{B 442}, 593
(1995).

[9] O. Dreyer, Phys. Rev. Lett. \textbf{90}, 081301(2003).

[10] G`t Hooft. Nucl. Phys. \textbf{B 256}, 727 (1985).

[11] A. Ghosh and P. Mitra, Phys. Rev. Lett. \textbf{73},
2521(1994).

[12] H. M. Lee and J. K. Kim, Phys. Rev. \textbf{D 54},
3904(1996).

\end{document}